\documentclass{sigchi}

\CopyrightYear{2018}
\doi{}
\isbn{}
\conferenceinfo{Full version of the L\@S'18 paper ``Measuring Item Similarity in Introductory Programming''}{}
\acmPrice{}

\usepackage{balance}       
\usepackage{graphics}      
\usepackage[T1]{fontenc}   
\usepackage{txfonts}
\usepackage{mathptmx}
\usepackage{color}
\usepackage{booktabs}
\usepackage{textcomp}

\usepackage{graphicx,booktabs}
\usepackage{amsmath}
\usepackage{microtype}        

\def\plaintitle{Measuring Item Similarity in Introductory Programming: Python
  and Robot Programming Case Studies}

\def\emptyauthor{}
\def\plainkeywords{programming;similarity;evaluation}

\makeatletter
\def\url@leostyle{%
  \@ifundefined{selectfont}{
    \def\UrlFont{\sf}
  }{
    \def\UrlFont{\small\bf\ttfamily}
  }}
\makeatother
\def\pprw{8.5in}
\def\pprh{11in}

\definecolor{linkColor}{RGB}{6,125,233}
\hypersetup{%
  pdftitle={\plaintitle},
  pdfauthor={\emptyauthor},
  pdfkeywords={\plainkeywords},
  pdfdisplaydoctitle=true, 
  bookmarksnumbered,
  pdfstartview={FitH},
  colorlinks,
  citecolor=black,
  filecolor=black,
  linkcolor=black,
  urlcolor=linkColor,
  breaklinks=true,
  hypertexnames=false
}

\usepackage{framed}
\newcommand{\arrow}[1]{\emph{Arrow~#1}}

\begin{document}

\title{\plaintitle}

\numberofauthors{5}


\author{%
  \alignauthor{Radek Pel\'anek\\
    \affaddr{Masaryk University}\\
    \affaddr{Brno, Czech Republic}\\
    \email{pelanek@fi.muni.cz}}\\
  \alignauthor{Tom\'a\v{s} Effenberger\\
    \affaddr{Masaryk University}\\
    \affaddr{Brno, Czech Republic}\\
     \email{tomas.effenberger@mail.muni.cz}}\\
  \alignauthor{Mat\v{e}j Van\v{e}k\\
    \affaddr{Masaryk University}\\
    \affaddr{Brno, Czech Republic}\\
     \email{vanekmatej@mail.muni.cz}}\\
  \alignauthor{Vojt\v{e}ch Sassmann\\
    \affaddr{Masaryk University}\\
    \affaddr{Brno, Czech Republic}\\
     \email{445320@mail.muni.cz}}\\
  \alignauthor{Dominik Gmiterko\\
    \affaddr{Masaryk University}\\
    \affaddr{Brno, Czech Republic}\\
     \email{ienze@mail.muni.cz}}\\
}


\maketitle

\begin{abstract}
  A personalized learning system needs a large pool of items for learners to
  solve. When working with a large pool of items, it is useful to measure the
  similarity of items. We outline a general approach to measuring the
  similarity of items and discuss specific measures for items used in
  introductory programming. Evaluation of quality of similarity measures is
  difficult. To this end, we propose an evaluation approach utilizing three
  levels of abstraction. We illustrate our approach to measuring similarity and
  provide evaluation using items from three diverse programming environments.
\end{abstract}



\section{Introduction}


A key part of learning is active solving of educational items (problems,
questions, assignments). For high-quality education we need large pools of
items to solve. Even teachers in classical education often work with many
items. In personalized computerized educational systems, the need for a large
pool of items is even higher -- if we want to provide a personalized experience
for individual learners, we need to be able to choose from a wide set of
items.

To use a large item pool efficiently, we need to be able to navigate it. For
this, it is very useful to be able to measure the similarity of individual
items. How can we measure the similarity of educational items? From a spectrum
of similarity measures, how do we pick a suitable one? These are the basic
questions that we address in this paper.

Similarity measures have many applications, particularly in adaptive learning
systems. Similarity measure can be very useful for automatic
\emph{recommendations} of activities. If a learner solved an item, but with a
significant effort, it may be useful to recommend as a next item another very
similar item, so that the learner can get more practice. On the other hand, if
a learner solved the current item easily, it is more meaningful to recommend
dissimilar item. If a learner struggles with an item, the system may provide as
a \emph{hint} a suitable worked example (a solution to a similar item) based on
the similarity measure. Similarity measure can be used in \emph{learner and
  domain modeling}: based on the similarity between items, we may define
knowledge components and estimate knowledge of learners. Similarity measures
may be also used in the \emph{user interface}, e.g., for enabling learners to
navigate the item pool and manually pick an item to solve, or for visualization
of the open learner model.

In addition to the use in automatic adaptation, similarity measures can be also
very useful for empowering humans by providing useful and actionable insight
(see~\cite{baker2016stupid} for a general discussion of this approach). For
developers of learning system and content creators, similarity measure
facilitates the management of an item pool, e.g., the identification of
redundant, duplicate, or missing items. Suitably presented data on item
similarity may be very useful for teachers, instructional designers, or
textbook authors. Such data may be useful for example for guiding the choice of
items for an exam -- typically we want items in an exam to be similar, but not
very similar to items practiced during learning. Data on item similarity may
provide impulses for the organization of classes, instructional materials, or
creation of other educational resources (e.g, worked examples). In systems with
crowdsourced content creation, the size of item pool may be very large and
similarity measures may be fundamental for efficient utilization of available
resources.

In most domains, there is no single correct measure of item similarity.
Particularly there may be a difference between item similarity based on
superficial features (a cover story) and deep features (a principle of
solution), which are related to different perspectives of novices and experts.
For some applications, it may be useful to work with several similarity
measures.

\begin{figure}[tb]

  \begin{framed}
    Write a function that outputs divisors of a given number.

    \smallskip

\begin{verbatim}
def divisors(n):
    for i in range(1, n + 1):
        if n % i == 0:
            print(i, end=" ")
    print()
\end{verbatim}
  \end{framed}
\medskip

\begin{framed}
  \begin{center}
    \includegraphics[width=.45\linewidth]{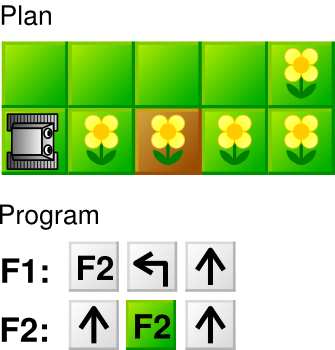}
  \end{center}
\end{framed}
\medskip

\begin{framed}
  \begin{center}
    \includegraphics[width=.7\linewidth]{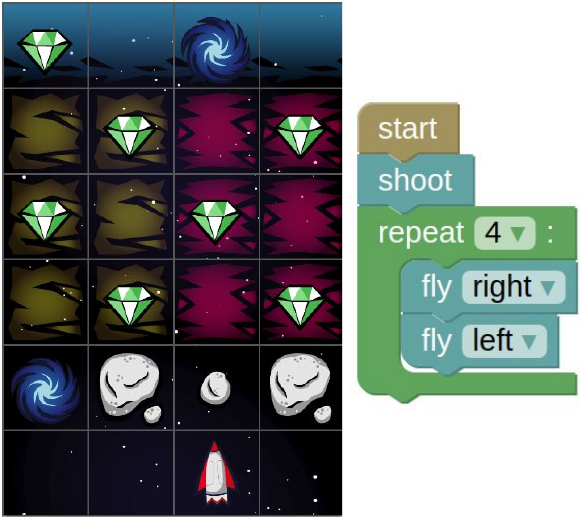}
  \end{center}
\end{framed}

\caption{Examples of programming problems with sample solutions from three
  programming environments (Python, Robotanist, RoboMission).}
  \label{fig:examples-problems}
\end{figure}

In this work we focus on the study of similarity of programming problems in the
context of introductory programming, specifically for programming exercises in
Python and programming problems with a robot on a grid, using a simplified
graphical programming language as used for example in popular \emph{Hour of
  code} activities. Figure~\ref{fig:examples-problems} provides examples of
problems and solutions from the three specific environments that we use
throughout this paper. There are other domains with complex items, where
measuring the similarity of items may be useful, for example mathematics,
physics, or chemistry. Introductory programming has the advantage that the item
statements and solutions are more easily processed and learners' solutions are
also readily available. Progress on a programming problem is done directly on a
computer and easily stored; as opposed to physics problems, which are still
more naturally solved on a paper. Currently, we focus only on programming
problems, but the general approach is applicable also to other settings --
therefore, we use the general term \emph{item} in our discussion.


\begin{figure*}[tb]
  \centering
  \includegraphics[width=.9\linewidth]{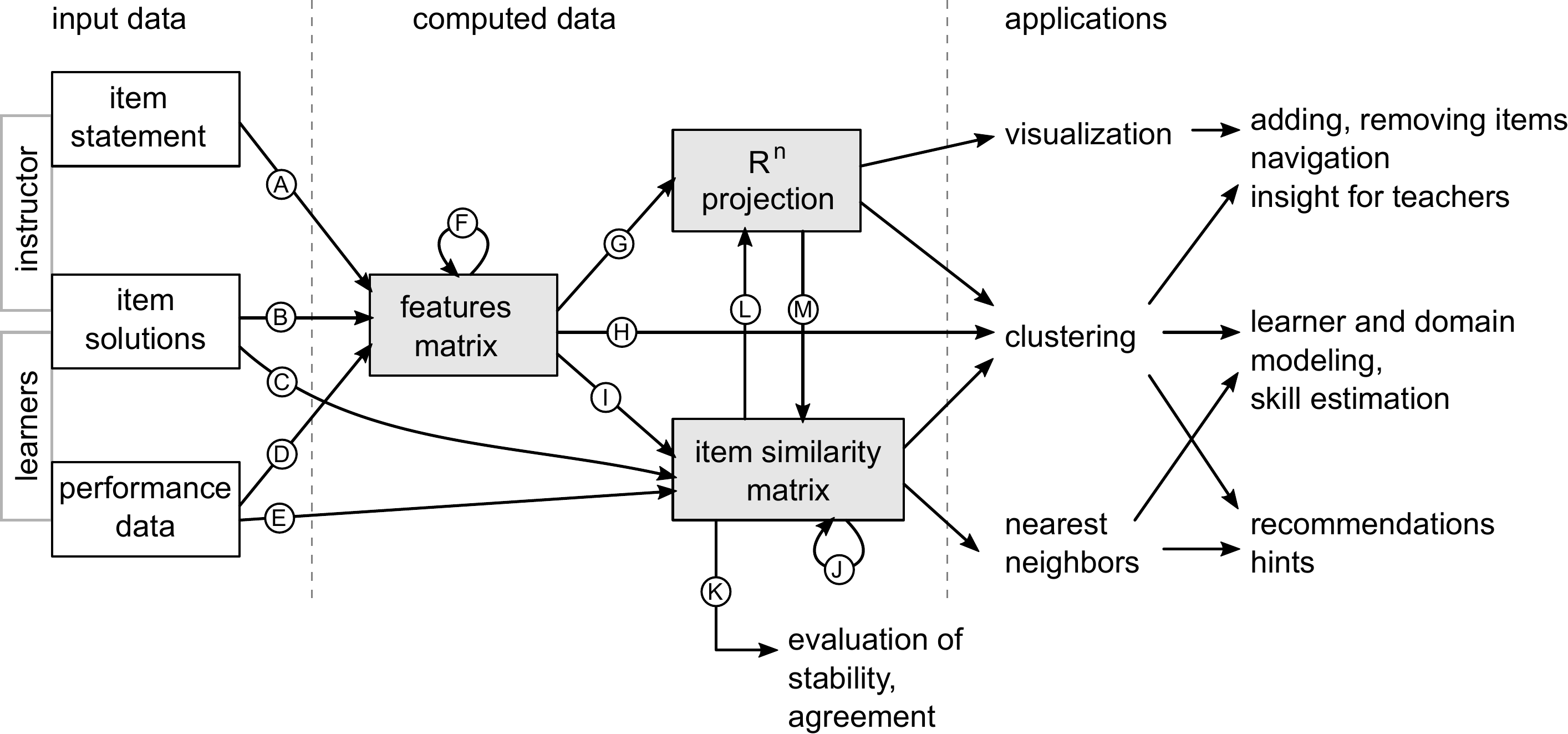}
  \caption{The general approach to computing and applying item similarity. The
    arrows that are discussed in more detail in the paper are denoted by
    letters and referenced in the text as \arrow{X}.}
  \label{fig:general-approach}
\end{figure*}

The general approach to measuring and using similarity of educational items is
outlined in Figure~\ref{fig:general-approach}: based on the available data we
compute similarity, which can be utilized in many ways. Several sources of data
can be used for measuring the similarity:
\begin{itemize}
\item an item statement: specification of the item that a learner should
  solve, e.g., as a natural language description of the task or an input-output
  specification,
\item item solutions: in the case of programming a solution to an item is a
  program written in a given programming language; we can use a sample solution
  provided by the item author or solutions submitted by learners,
\item data about learners' performance: for example item solving times, number
  of attempts needed, or hints taken.
\end{itemize}


In analyzing and applying similarity it is useful to explicitly distinguish two
matrices, which naturally occur in computations:
\begin{itemize}
\item a \emph{feature matrix}, in which rows correspond to items and columns to
  features of items (e.g., keywords occurring in an item statement or an item
  solution),
\item an \emph{item similarity matrix}, which is a square matrix $S$, where
  $S_{ij}$ denotes similarity of items $i$ and $j$.
\end{itemize}

Figure~\ref{fig:general-approach} shows typical steps in the computation and
application of similarity. For each step there are many possible choices for
their specific realization. For example, the \arrow{I} (computing a similarity
matrix from a feature matrix) can be done using Euclidean distance, Pearson
correlation coefficient, cosine similarity, and many other measures. Similarly,
there are many specific ways how to transform an item solution into a feature
matrix (\arrow{B}) and many algorithms for performing clustering (\arrow{H}).
Moreover, individual steps are independent and can be combined.

In this work we focus on measuring similarity, i.e., constructing the item
similarity matrix. We do not discuss in detail different applications, since it
is necessary at first to properly clarify how to compute similarity. Our main
contributions are the following. We provide overview and terminology for the
problem of ``measuring the similarity of programming items'' and a systematic
mapping of available choices and approaches to the problem. As opposed to
related work, which typically utilizes only a single setting, we explore and
evaluate different approaches to computing similarity in the context of three
different programming environments (Python programming and two robot
programming environments). We systematically analyze the role of different
choices in the computation in item similarity, utilizing analysis with three
levels of abstraction.



\section{Related Work}

The overall approach outlined is Figure~\ref{fig:general-approach} is related
to the distinction between pairwise data clustering versus feature vector
clustering, which has been studied in the general machine learning
research~\cite{hofmann1997pairwise,roth2003optimal}.

A specific domain in which item similarities and clusters have been extensively
explored are recommender systems, e.g., in neighborhood-based
methods~\cite{desrosiers2011comprehensive}, item-item collaborative filtering
\cite{deshpande2004item}, and content based techniques~\cite{lops2011content}.
Similarity measures has been used for clustering of
items~\cite{o1999clustering,park2008long} and also for clustering of
users~\cite{sarwar2002recommender}.

In the domain of educational data mining, previous research explored similarity
based on performance data. In~\cite{edm17-similarity}, authors study similarity
of items and focus on comparison of different similarity measures. In
\cite{kaser2013cluster}, authors study similarity and clustering of users and
provide detailed description of processing pipelines for computing similarity.
Another related line of research in educational data mining is based on the
concept of a Q-matrix~\cite{tatsuoka2005rule,barnes2005q}. A Q-matrix can be
seen as a generalization of a clustering; it provides mapping between items and
concepts, one item can belong to several concepts. The Q-matrix can be
constructed or refined based on the performance
data~\cite{desmarais2014refinement}.

In the context of programming, previous research have studied similarity of
programs for a single programming problem. This has been done in two different
settings. At first, in plagiarism detection~\cite{lancaster2004comparison},
where the goal is to uncover programs that have been copied (and slightly
obfuscated). At second, in the analysis of submissions in massive open online
courses, where a single assignment can have a very large number of submissions.
The goal of the analysis in this context is to get insight into the learning
process or multiply instructor leverage by propagating feedback
\cite{piech2012modeling,nguyen2014codewebs,yin2015clustering}, or to
automatically generate hints or examples based on learners' submissions
\cite{rivers2017data,iii2014generating,gross2014select}. These works often use
computation of similarity between programs as one of the steps, employing
techniques like \emph{bag-of-words} program representation, analysis of
abstract syntax trees, and tree edit distance. We employ these techniques for
our analysis.




The most closely related work is by Hosseini et al.~\cite{hosseini2017study},
who consider specifically the question of similarity of programming items or
items and worked out examples, using introductory programming problems
concerning the Java language. They, however, use only items of the type ``What
is the output of a given program?'', whereas we consider more typical
programming problems where learners are required to write a program.



\section{Items in Introductory Programming}
\label{sec:items-intr-progr}

Throughout the paper we use the general notation of educational items, since
many aspects of our work are generally applicable. Our focus is, nevertheless,
on programming problems. To further specify the context of our research, we now
describe examples of specific introductory programming environments and
programming items. These environments and items are also used in evaluation.

\subsection{Programming in Python}

The standard item in introductory programming is to write a program
implementing a specified functionality in a general purpose programming
language. Examples of such items are ``Write a function that outputs divisors
of a given number.'', ``Write a function that outputs ASCII art chessboard of
size N.'', ``Write a function that replaces every second letter in a string by
X.'', or ``Write a function that outputs frequencies of letters in a given
text.''.

An item statement in this context is given by a natural language description of
the task (as illustrated above), usually with some examples of input-output
behavior. An item typically contains also a sample solution, which can be used
for automatic evaluation of learners solutions.

For our analysis we consider programming problems in Python, which is a typical
language used for introductory programming. We use 28 items from the system
\texttt{tutor.fi.muni.cz}, for these items we have data about learner
performance (time taken to solve the problem). We also use 72 items for which
we have only the item statement -- these items are used in a large university
programming class, but without data collection.


\subsection{Robot Programming}

A less standard, but very popular approach to teaching basic programming
concepts is to use a simplified programming environment to program robots on a
grid. Problems of this type are often used in the well-known \emph{Hour of
  code} activity, with participation of millions of learners.

An item in this context is given by a specific world, which is typically a
grid with a robot and some obstacles or enemies, and some restrictions placed
on a program, e.g., there is often a limit on the number of commands, so that
the problem has to be solved using loops instead of long list of simple
commands. Solutions are written in a restricted programming language, which
contains elementary commands for a robot (move forward, turn, shoot) and basic
control flow commands. The program is often specified using graphical
interface, a common choice today is the Blockly interface~\cite{fraser2015ten}.

For our analysis we use two environments of this type (illustrated in
Figure~\ref{fig:examples-problems}). \emph{Robotanist} (available at the web
\texttt{tutor.fi.muni.cz}) uses very simple programming commands (arrows for
movement, colors for conditions), but the solution of problems is often
nontrivial due to limits on the number of commands. Solutions thus require
careful decomposition into functions and use of recursion. For this environment
we have 74 items, including data on learners' performance (over 110 thousand
problem solving times). \emph{RoboMission} (\texttt{en.robomise.cz}) uses a
richer world (e.g., meteorites, worm holes, colors, diamonds) and is programmed
in the Blockly interface using a richer set of commands (movement, shooting,
several types of conditions, repeat and while loops). For this environment we
have 75 items. Data on learners' performance are available, but are not yet
sufficiently large, therefore we use only the item statement and sample
solutions for our analysis.



\section{Computing Similarity}

We structure the discussion of similarity computation into two steps. At first,
we cover the basic processing of input data to get a feature matrix or a
similarity matrix. At second, we consider transformations of the computed
matrices.

In our discussion we use similarity measures (higher values correspond to
higher similarity). Related work sometimes considers distance (dissimilarity)
measures (lower values correspond to higher similarity). This is just a
technical issue, as we can easily transform similarity into dissimilarity by
subtraction.


\subsection{Processing the Input Data}

In this step we need to get from the raw input data to the matrix form. The
details of these steps depend on the type of data and specifics of the used
items -- the input data may include natural language text, programs written in
a programming language, formal description of the robot world, etc.

\subsubsection{Item Statement}

To compute similarity based on an item statement, it is natural to go through
the feature matrix (\arrow{A}). The choice of features depends on the specific
type of programming environment.

For the standard programming items where the item is specified using a natural
language, the features have to be obtained from this text. A basic technique is
to use the bag-of-words model, i.e., representing the text by a vector with the
number of occurrences of each word (with standard processing, e.g.,
lemmatization and omission of stop words). Other features may be derived from
the input-output specification, e.g., features describing types of variables
(e.g, integer, string, list).

In the case of robot programming, an item is typically specified by a robot
world description and a set available commands. Natural language description
may be present, but typically does not contain fundamental information. Basic
features thus correspond to the aspects of the word and the used programming
language, e.g., in the RoboMission setting these are word concepts like
diamond, worm hole, meteorite and programming aspects like the
limit on the number of commands.

\subsubsection{Item Solution}

Another approach to measuring similarity of items is to utilize solutions,
i.e., programming codes that solve an item. The basic approach is to utilize a
single solution -- either the sample solution provided by the item author, or
the most common learner solution. Here we have two natural approaches to
computing similarity: via feature matrix or by direct computation of
similarities.

With the \emph{features approach} (\arrow{B}) we analyze the source code
(typically using traversal of the abstract syntax tree) to compute features
describing occurrence of programming concepts and keywords like while, if,
return, print, or use of operators. The basic approach is again the
bag-of-words model, only applied to programming keywords instead of words in a
natural language. In this representation we lose information about the
structure of the program -- we only retain the presence of concepts and the
frequency of their occurrence. In addition to features corresponding to
keywords, we can use features like the use of functions, the nested depth, or
the length of a code. In some settings it may also be possible to use features
based on input-output behavior (e.g., type of the input and output).

The second approach is to compute directly item similarity matrix (\arrow{C})
by computing \emph{edit distance} between the selected solutions for the two
items. The edit distance can be computed in several ways:
\begin{itemize}
\item tree edit distance~\cite{bille2005survey} for the abstract syntax tree,
  potentially with some specific modifications for programs (in the specific
  programming language),
\item the basic Levenshtein edit distance for the canonized code, which is
  applicable particularly for the robot programming exercises, where programs
  can be relatively easily canonized,
\item edit distance applied to the sequence of actions performed (API
  calls), this is again easily applicable particularly for the robot
  programming exercises, which have a clear API; \cite{piech2012modeling}
  uses this approach together with the Needleman-Wunsch global DNA alignment
  for measuring edit distance.
\end{itemize}

So far, we have considered only a single solution. Typically, we have more
solutions -- programming problems can be solved in several ways, so it may be
useful to have multiple sample solutions. We can also collect learners'
solutions and use them for analyzing item similarity. The basic approach to
exploiting multiple solutions is to compute the feature matrix for each of them
and then use a (weighted) average of matrices. There are, however, other
possible choices, particularly in the direct computation of item similarity
(\arrow{C}) based on edit distance it may make sense to use \emph{minimum}
rather than \emph{average}.

\subsubsection{Performance Data}

Finally, we may use data on performance of learners while solving items, e.g.,
the correctness of their solutions, the number of attempts, problem solving
time, or hints taken.

These data can be transformed into features (\arrow{D}) like average
performance, variance of performance, or ratio of learners who successfully
finish an item. Such features in most applications will not carry sufficiently
diverse information to compute useful similarity between items, but these
features may be useful as an addition to feature matrix based on an item
statement or solution.

We can, however, compute similarity directly from the performance data
(\arrow{E}): similarity of items $i$ and $j$ is based on the correlation of
performance of learners on items $i$ and $j$ with respect to a specific
performance measure (the correlation is computed over learners who solved both
items $i$ and $j$). This approach has been previously thoroughly evaluated in
the case of binary (correctness) performance data~\cite{edm17-similarity}. In
the case of programming problems, it is natural to use primarily problem
solving times (rather than correctness).



\subsection{Data Transformations}

Once we compute the item features or basic item similarities, we can process
them using a number of transformations. In contrast to the above described
processing of input data, which necessarily involves details specific for a
particular type of items, the data transformation steps are rather general --
they can be used for arbitrary feature matrices and are covered by general
machine learning techniques. What may be specific for programming or for
particular input data is the choice of suitable transformations.

\subsubsection{Feature Transformations and Combinations}

The basic feature matrix obtained by data processing contains for each feature
raw counts, e.g., the number of occurrences of a keyword in a sample
program. Before computing similarity it is useful to normalize the values by
transforming values in the feature matrix (\arrow{F}), i.e., by performing
transformations that take the feature matrix and produce new, modified feature
matrix.

Examples of simple transformations are \emph{binarization} (very coarse grained
normalization), \emph{normalization by dividing by a maximal value} for each
feature to get values into the $[0, 1]$ interval, or \emph{log transform} (to
limit the influence of outlier values). A typical transformation, particularly
in the context of the bag-of-words features, is the \emph{TF-IDF} (term
frequency--inverse document frequency) transformation.

Often we can obtain several feature matrices (or item similarity matrices)
corresponding to different data sources or multiple solutions. We can combine
these matrices in different ways, the basic ones being: \emph{average}
(assuming additive influence of data sources), \emph{min} (assuming conjunctive
influence of data sources), \emph{max} (assuming disjunctive influence of data
sources).


\subsubsection{Computing Similarity}

When we compute the similarity matrix based on the feature matrix (\arrow{I})
or on its projection into $R^n$ (\arrow{M}), we have a vector of real values
for each item; the similarity of a tuple of items is computed as the similarity
of their vectors. This is a common operation in machine learning, with many
choices available. The common choices are cosine similarity, Pearson
correlation coefficient, and Euclidean distance (transformed into similarity
measure by subtraction). These measures are used widely in recommender systems,
with the experience that the choice of suitable measure depends on a particular
data set~\cite{desrosiers2011comprehensive}.

The suitable choice of a similarity measure depends also on the steps used to
compute the feature matrix and on the purpose of computing similarity. As an
example, consider two programming problems, where solutions use the same
concepts (keywords), but one of the solutions is longer and uses the keywords
multiple times. If we use normalization, the feature vectors will be (nearly)
the same and the items will end up as very similar for any similarity measure.
If we do not use normalization, the items will end up as very similar when we
use cosine similarity and correlation coefficient, but as different when we use
Euclidean distance. We cannot give a simple verdict, which one of these is
better, since this may depend on the intended application.

\subsubsection{Projections}

From feature matrix or item similarity matrix we can compute projection to
$R^n$ (\arrow{G} and \arrow{L}). Such projection is typically used for
application, particularly for visualization of items. It can, however, also be
a useful processing step in the computation of item similarities, for example
in the case of correlated features we can use the principal component analysis
(PCA) for decorrelating features (\arrow{G}) and then compute similarities
based on the principal components (\arrow{M}).

There are many techniques for computing low dimensional projections, for
feature matrix (\arrow{G}) the popular choices include the basic linear PCA and
the nonlinear t-SNE~\cite{maaten2008visualizing}, for similarity data
(\arrow{L}) the basic technique is the (non-metric) multidimensional scaling.




\section{Evaluation}

As the previous section shows, there is a wide number of techniques that can be
used for computation of similarity. Moreover, individual steps can be combined
in many ways. What is a good approach to computing similarity? Which decisions
matter?

Evaluation of similarity measures is difficult, because there is no clear
criterion of quality of measures. The suitability of measure depends on a
particular way it is used in a specific application. However, it is very useful
to get an insight into similarity measures in application independent way. To
this end we analyze similarity measures at several levels of abstraction:
\begin{enumerate}
\item Similarity of items for a specific similarity measure. This allows us to
  get the basic understanding of what kind of output we can obtain.
\item Agreement of different measures (across all items). This allows us to get
  understanding how much do the choices made in the computation matter.
\item Analysis of different agreement measures. To measure agreement between
  similarity measures we must also choose some method of quantification
  (\arrow{K}). There are several natural candidates. Does it matter, which
  one we choose?
\end{enumerate}
Finally, as an illustration of an application specific evaluation, we consider
evaluation of measures with respect to clustering of items.

We perform our analysis across different contexts -- different programming
environments and sets of items. We use the settings described in
Section~\ref{sec:items-intr-progr}, i.e., Python, Robotanist, RoboMission. This
gives us an insight into how the observed results generalize.


\subsection{Similarity of Items for a Specific Measure}

We start with the basic kind of analysis -- exploration of results obtained
using a specific similarity measure; i.e., we pick a specific way to compute
item similarity matrix and then explore the matrix, its visualization and
projection.

\begin{figure}[tbh]
  \centering
  \begin{tabular}{cc}
  \includegraphics[width=\linewidth]{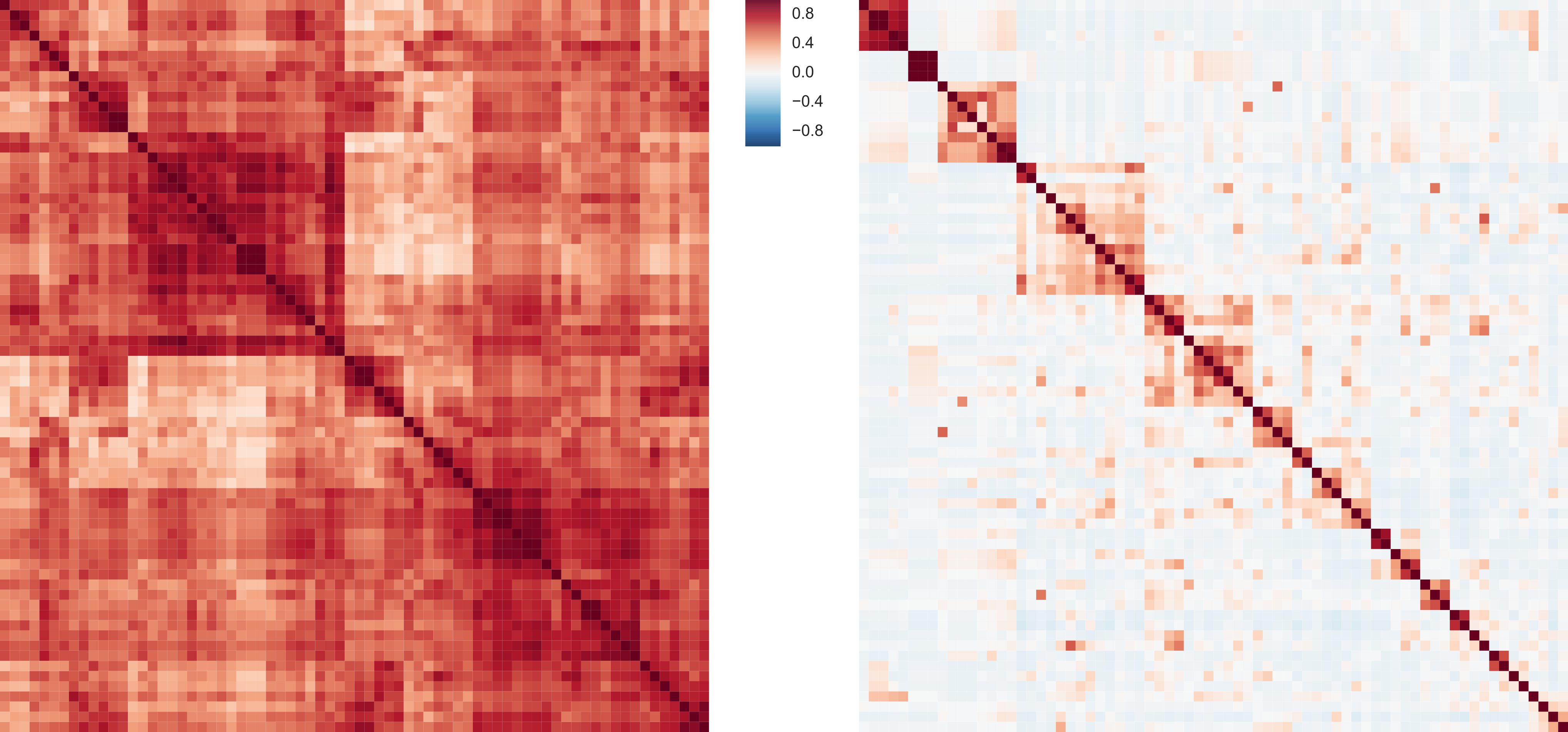}
  \end{tabular}
  \caption{Item similarity matrices for 72 Python programming problems. Left:
    similarity computed using features based on keywords in problem sample
    solutions, logarithmic transformation, and correlation. Right: similarity
    computed using features based on words in natural language item statement,
    TF-IDF transformation, and correlation. Note that items are ordered by
    hierarchical clustering and although the matrices show same items, each
    uses different ordering.}
  \label{fig:similarity-examples}
\end{figure}

Figure~\ref{fig:similarity-examples} shows two similarity matrices for 72
Python programming problems. The matrix based on sample problem solutions is
dense -- many keywords (e.g., \texttt{print}, \texttt{for}) are shared by many
items. The matrix based on item statements is sparse -- in this case we are
using very brief item statements (typically one sentence), and thus items share
words only with several other items.


\begin{figure}[tbh]
  \centering
  \includegraphics[width=0.6\linewidth]{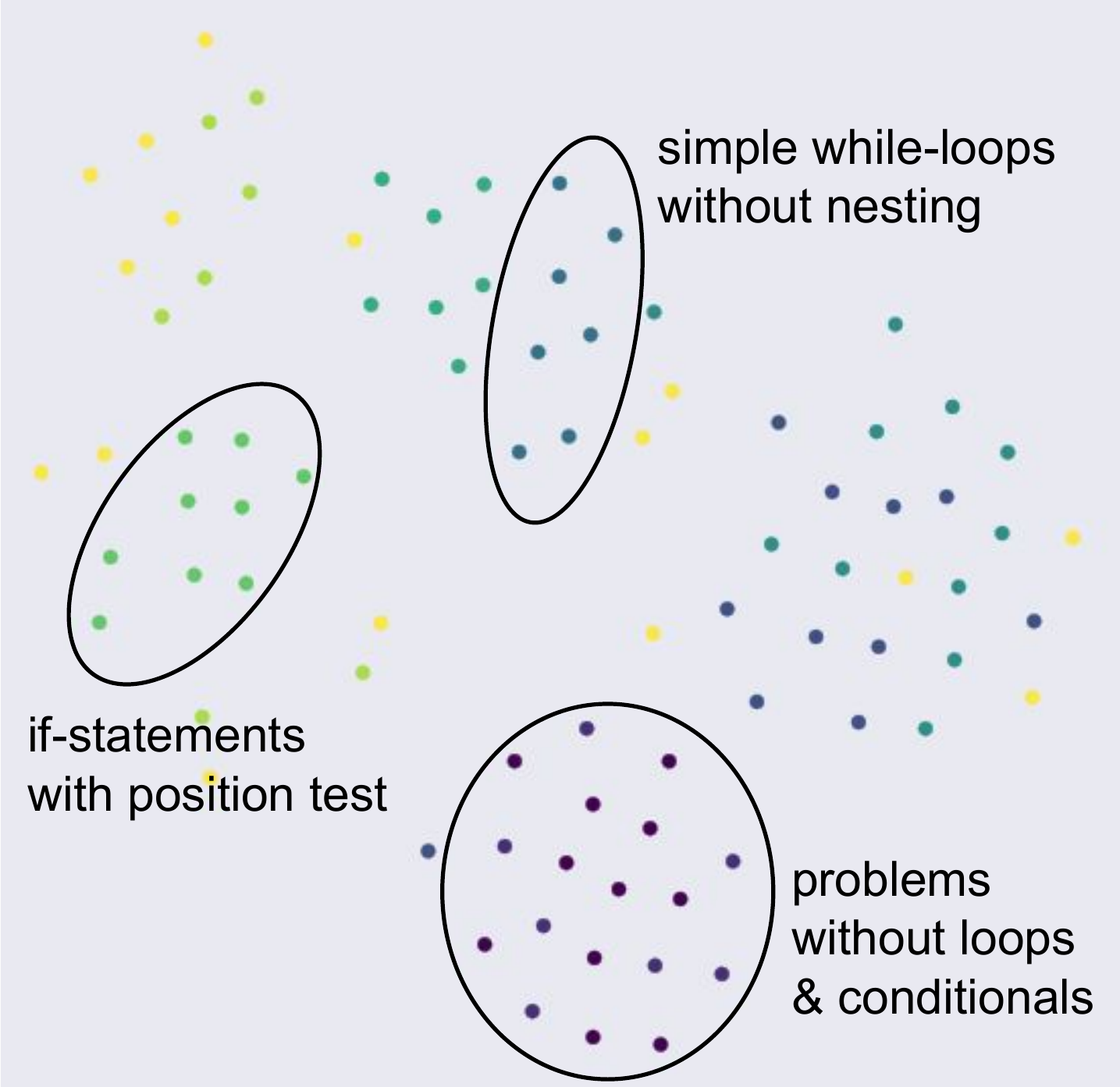}
  \caption{Projection into plane of 86 items from RoboMission via tSNE.
    Items are coloured by their manual division into 9 levels, which is
    currently used in the production system (the brighter the colour, the
    more difficult level -- the easiest items are violet, the most difficult
    yellow). The projection leads to meaningful groups of items, three of them
    are highlighted. }
  \label{fig:projection-example}
\end{figure}

Figure~\ref{fig:projection-example} shows a projection into a plane of a feature
matrix for items from RoboMission. Features are based on both item statement and
sample solution (bag-of-words), with log and IDF transformation, and divided by
the maximum value. Because statement features have significantly higher counts,
these transformations are important in order for the projection to be influenced
by both the statement and the solution.
Without these transformations, items from different levels end up
noticeably more mixed up in the projection.


The usefulness of measures based on the performance data clearly depends on the
size of available data -- we need sufficiently large data so that the measures
are stable. As a basic check of stability we split the performance data into
two independent halves, compute the measure for each half, and then check their
agreement. We performed this evaluation for our Python programming data and the
Robotanist problem, where we have problem solving times available. The
resulting correlation is 0.54 (27 Python problems) and 0.28 (74 Robotanist
problems) -- high enough to conclude that there is a significant underlying
signal in the performance data (and not just a random noise), but clearly the
size of data is not yet sufficiently large to provide stable similarity measure
for practical applications. We experimented with restricting the data to only
items which were solved by large number of learners. When we use items with at
least 400 solutions, the measures are getting stable (correlation over 0.8 in
both cases), but we have data of this size only for about one half of our items.



\subsection{Agreement between Measures}

Our main goals in this paper are related to this level of analysis -- dealing
with questions like ``What is the relation between different similarity
measures?'' and ``Which steps in the pipeline are most important?''.

\begin{figure}[tbh]
  \centering
  \includegraphics[width=\linewidth]{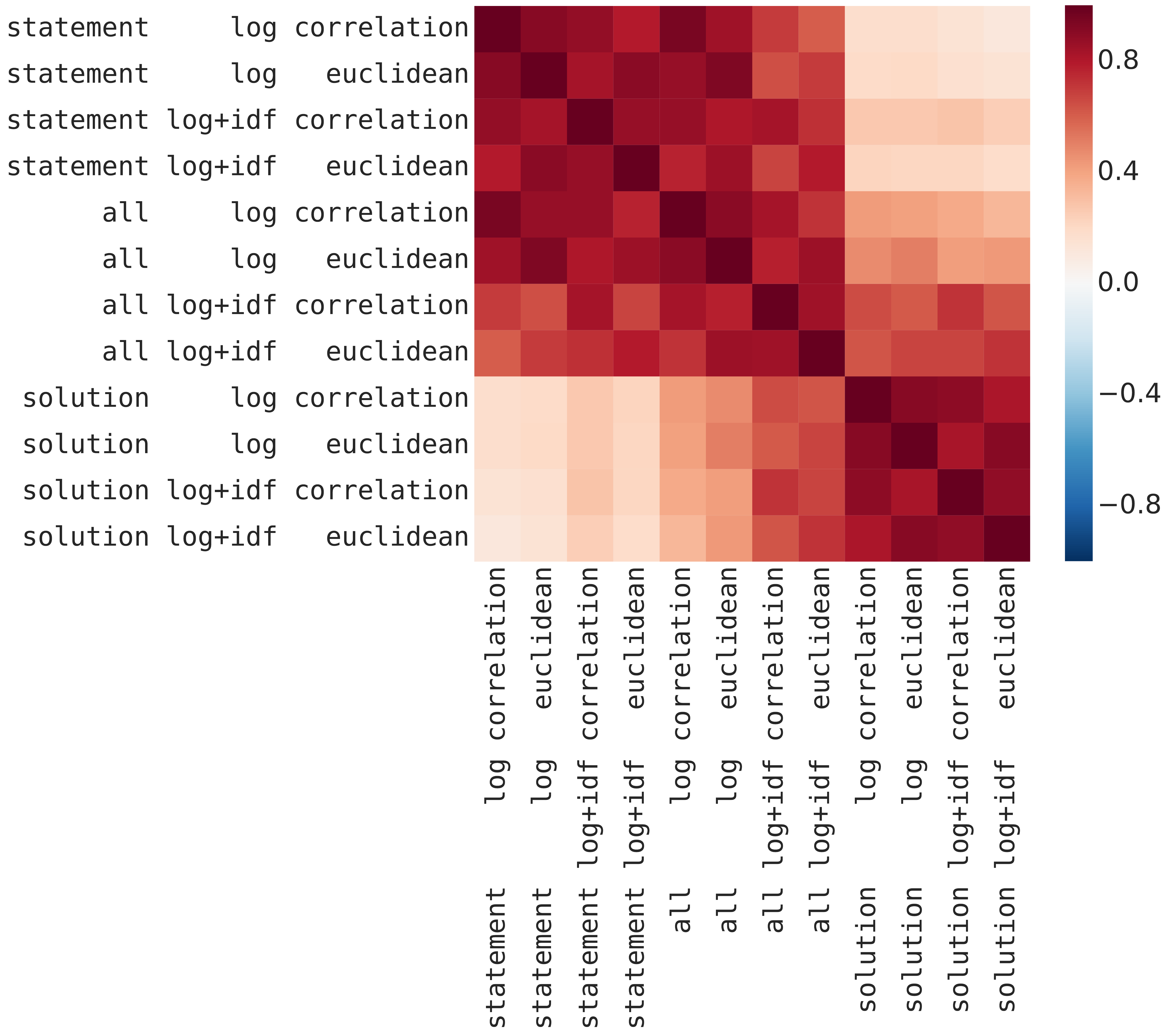}
  \caption{Agreement between 12 similarity measures for problems in
    RoboMission. Bag-of-words features from either problem statement, sample
    solution, or both, transformation either log, or log+IDF, similarity
    function either correlation, or (subtracted) Euclidean distance.}
  \label{fig:robomise-corr-measures2}
\end{figure}

\begin{figure}[tbh]
  \centering
  \includegraphics[width=.8\linewidth]{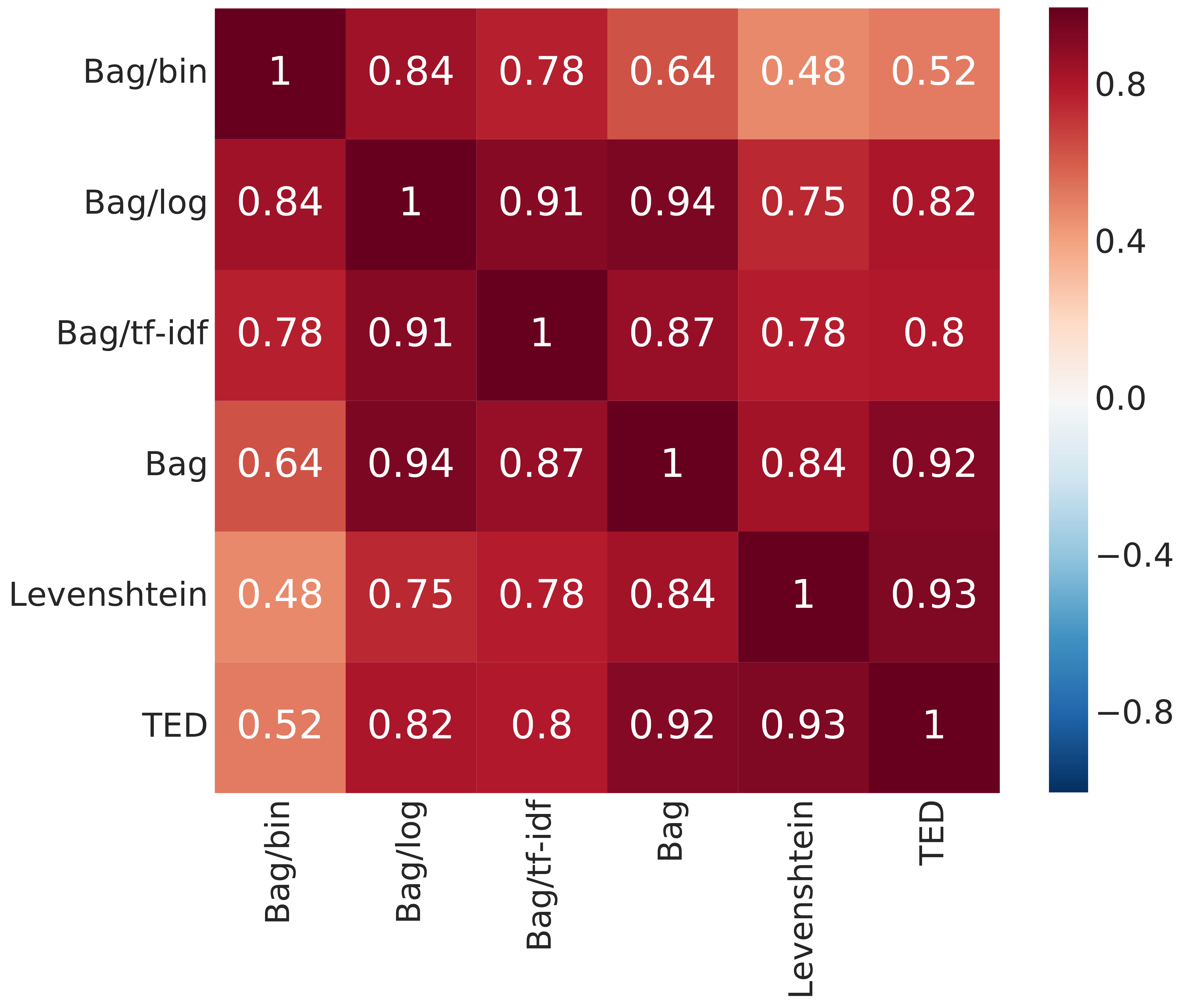}
  \caption{Agreement between similarity measures that use sample solutions (for
    problems in RoboMission), ordered from the most structure-ignoring approach
    on the top/left (bag-of-words, binarization transformation, Euclidean
    distance) to the most structure-based approach on the bottom/right (Tree
    Edit Distance).}
  \label{fig:robomise-corr-measures3}
\end{figure}

To analyze agreement between two similarity measures, we first compute the item
similarity matrix for of each of them (obtaining two matrices of the type
displayed in Figure~\ref{fig:similarity-examples}) and then compute the
agreement as a correlation of values in these two similarity matrices. For a
set of similarity measures, this gives us a matrix of agreement values, as
illustrated in Figure~\ref{fig:robomise-corr-measures2} and
Figure~\ref{fig:robomise-corr-measures3}. These figures are for the RoboMission
environment, but they illustrate trends that we see across all our data sets.

Similarity measures based on item statement vs. solution are only weakly
related; they focus on different aspects of items and their similarity (see
Figure~\ref{fig:robomise-corr-measures2}). However, the relationship can be
stronger, if the item statements include more details or constraints about the
solution, such as a set of allowed programming blocks that the learners can use
to build their solution.


Measures that use the same source of data only in different ways are typically
highly correlated. Figure~\ref{fig:robomise-corr-measures3} shows agreement
between measures that use sample solutions. Low agreement is only between the
binarized bag-of-words features, which completely ignores the structure of the
solution, and Levenshtein or Edit Tree Distance approaches, which are on the
opposite spectrum of the focus on the structure. Using a bag-of-words with
log-counts (possibly with some feature weights normalization, such as IDF) is a
reasonable compromise. The effect of the choice of a function for computing
similarity from feature matrix (\arrow{I}) seems to be small -- in
Figure~\ref{fig:robomise-corr-measures2} we see comparison of correlation and
Euclidean distance.

For measures that combine multiple sources of data, feature normalization is
important to adjust for different scales and avoid one source of data being far
more important than the other. This issue can be seen in
Figure~\ref{fig:robomise-corr-measures2} -- without IDF normalization, the
measures that use all features are correlated to those that use item statement,
but not to those that use the solution.

\begin{figure}[tbh]
  \centering
  \includegraphics[width=.95\linewidth]{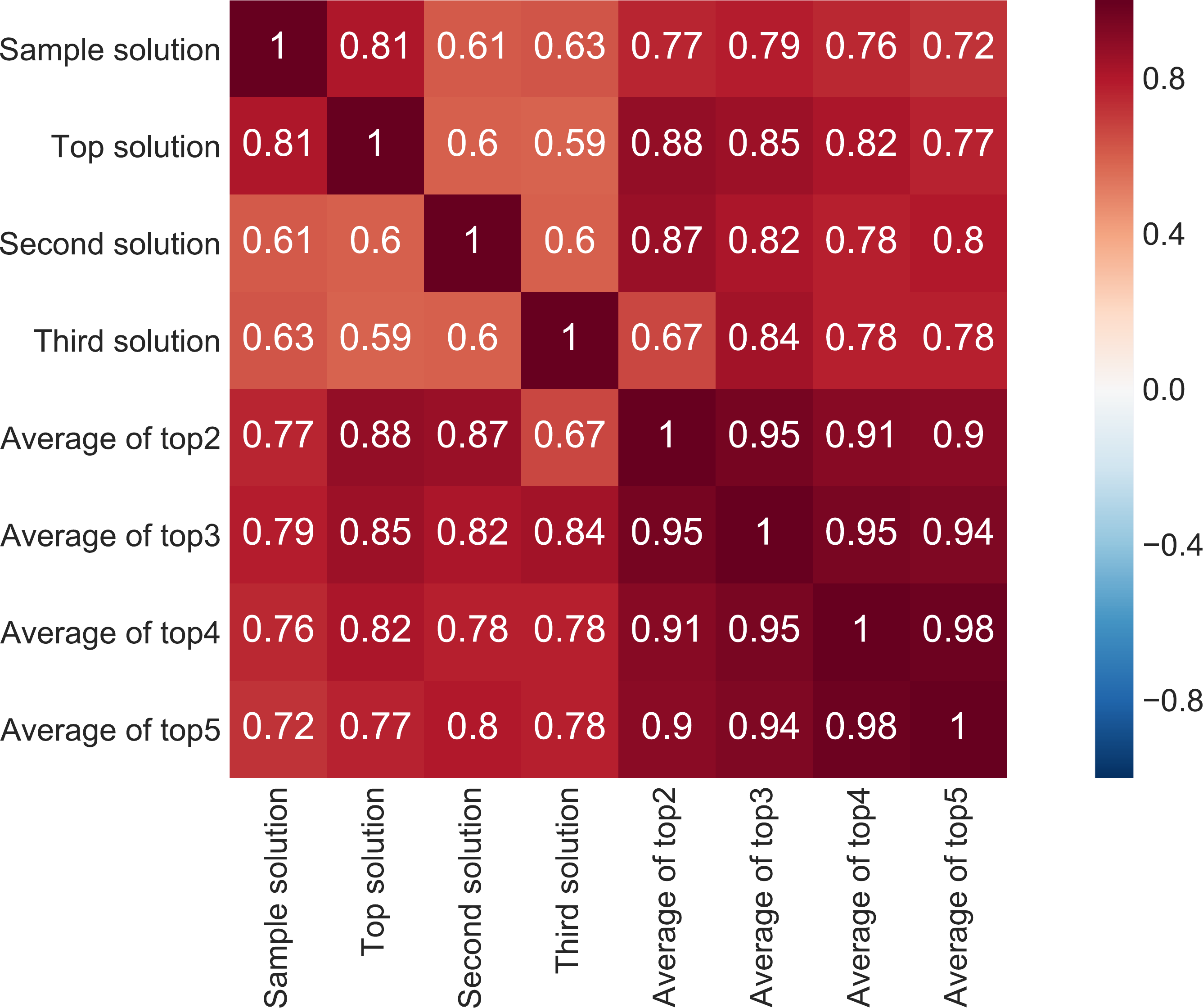}
  \caption{Agreement between similarity measures that use sample solution and
    (multiple) learners' solutions (for Python programming).}
  \label{fig:corr-measures}
\end{figure}

Figure~\ref{fig:corr-measures} shows relations between measures that utilize
different types of solutions. Here we use data on Python programming, where
learners solutions are available, and compare measures of similarity based on
the sample solution (provided by the author of the item), the top most common
learners' solutions, and the averages of several top solutions. The measures
based on sample solution and top learner solution have high overall
correlation, but for individual items they may differ quite significantly -- in
some cases the sample solution and top learner solution may be very different
as even simple programming problems like computing factorial can be written in
widely different ways: using for loop, while loop, or recursion. When data on
learner solutions are available, using the average of the 3 most common
solutions seems to be good, robust approach.

For the Robotanist and Python programming, where we have data on learners'
performance, the correlation between measures based on solution and measures
based on performance is between 0.2 and 0.4. This is a weak agreement, but
since the measures based on performance are not yet stable, even this weak
agreement may indicate a relationship between these different approaches to
measuring similarity. This relationship needs to be explored for larger data
sets.


\subsection{How to Measure Agreement?}

In the previous analysis we used simple correlation to measure agreement, i.e.,
to quantify agreement of two similarity matrices, we flatten them into vectors
and then compute Pearson's correlation coefficient over these vectors. But there
are other choices to measuring agreement. Particularly in applications, where
the intended use of similarity measures is to pick the closest neighbors (e.g.,
recommendation of similar exercises), it may be more relevant to measure
the agreement of rankings or agreement on top $N$ positions.

Do the results presented in the previous section depend on our choice of the
approach to measuring agreement? To explore this question we went with our
analysis up one level of abstraction. Now we analyze the relationship between
different measures of agreement of similarity measures. Specifically, we
compare measures of agreement based on correlation of flatten matrices and
measures of agreement based on top $N$ positions. When measuring agreement of
two similarity matrices $S_1$ and $S_2$ with respect to top $N$ positions, we
use the following approach: for each item we find $N$ most similar items with
respect to $S_1$ and $S_2$; we compute the size of intersection of these two
sets; finally, we average over all items and normalize by $N$.

\begin{figure}[tbh]
  \centering
  \includegraphics[width=.7\linewidth]{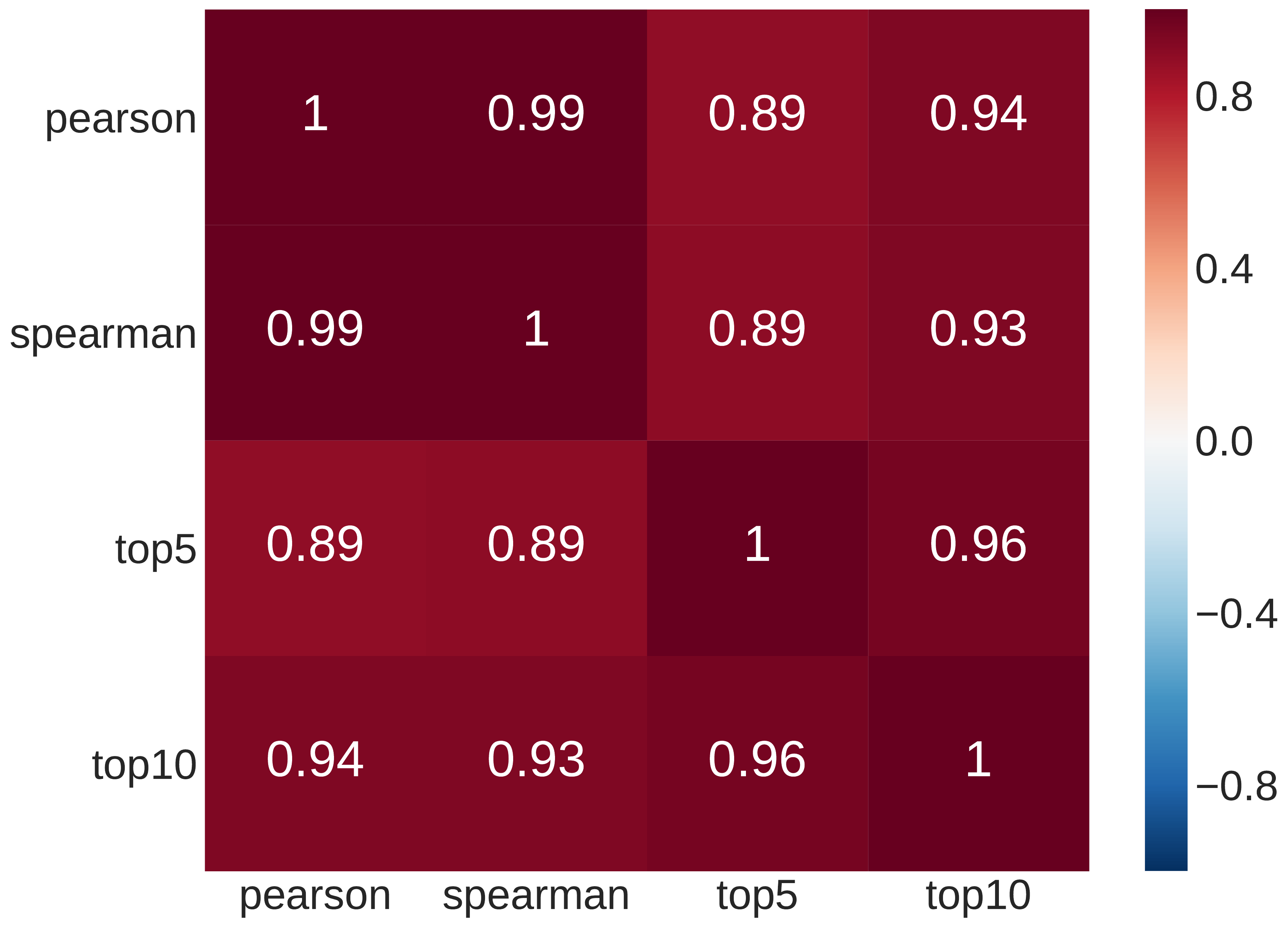}
  \caption{Correlation between methods for evaluating agreement: RoboMission.}
  \label{fig:corr-agreement}
\end{figure}

Evaluation works in this way: we pick several similarity measures, for each of
them we compute similarity matrix (as in Figure~\ref{fig:similarity-examples});
then for each measure of agreement we compute the agreement matrix of all the
chosen similarity measures (as in Figure~\ref{fig:robomise-corr-measures2});
finally, we compare these agreement matrices.  Example of this analysis is in
Figure~\ref{fig:corr-agreement}. There is some difference between using
correlation and ``Top 5'' measure of agreement, but this difference should not
have large impact on conclusions about the choice of similarity measures. Note
that for this analysis we use the basic correlation, which is again an ad hoc
choice, but for now we have decided to stop our analysis here and not to pursue
other levels of abstraction.

\subsection{Impact on Measure Application}

Our focus in this paper is on evaluation of different similarity measures
in a general scenario. However, the evaluation can also target on a specific
application for the similarity measure. In this section, we present an example
of such analysis.

In the production system, items in RoboMission are divided into 9 levels, which
were specified manually by the system developers. We examine how well we are
able to reconstruct this labeling using different similarity measures. For each
similarity measure, we compute the similarity matrix and then use k-means
clustering for this matrix with $k$ set to 9 (the number of levels). The
obtained clusters are compared with the manual labeling using the Rand Index.

The results (averaged over 10 runs) are shown in
Table~\ref{tbl:kmeans-results}. As the manually created levels differ mainly by
used programming concepts, such as loops and conditionals, it helps to multiply
solution features to give them higher weight. However, if the item statements
were more important for the division into levels, then the statement features
should have the higher weights instead. The weights can be even learned to
optimize an objective function for given application, provided there is enough
data to avoid overfitting.

\begin{table}[tbh]
\centering

\caption{Agreement between manual labeling and clusterings obtained by
  k-means algorithm for similarity matrices for different similarity
  measures. Transformations used with bag-of-words features: log, dividing by
  maximum value, multiplying by IDF, multiplying all solution features by the
  factor of 5.}
\label{tbl:kmeans-results}

\medskip

\begin{tabular}{l l}
Similarity Measure & Rand Index \\
\midrule
Bag / log+max+idf+weights / correlation & 0.49 \\
Bag / log+max+idf+weights / euclidean   & 0.39 \\
TED                                     & 0.30 \\
Bag / log+max+idf / euclidean           & 0.29 \\
Bag / bin / euclidean                   & 0.28 \\
Bag / log+max / euclidean               & 0.28 \\
Bag / bin / correlation                 & 0.27 \\
Levenshtein                             & 0.26 \\
Bag / log+max+idf / correlation         & 0.26 \\
Bag / log+max / correlation             & 0.22 \\
Bag / log / euclidean                   & 0.18 \\
Bag / log / correlation                 & 0.14 \\
Bag / no transformation / euclidean     & 0.11 \\
Bag / no transformation / correlation   & 0.09 \\
\end{tabular}
\end{table}




\section{Discussion}

We propose a systematic approach to defining and analyzing item similarity. We
discuss the issue specifically in the context of introductory programming, but
the approach is general and can be applied also for other complex items, e.g.,
in mathematics of physics.

\subsection{Systematic Approach to Similarity Measures}

Measuring similarity of items is a complex problem, because it can be tackled
in many ways (particularly for complex items like programming problems) and it
is hard to evaluate the quality of measures. We propose a systematic approach
to studying similarity measures (outlined in
Figure~\ref{fig:general-approach}), which makes explicit the many choices that
we need to make to specify a similarity measure.

We also propose a systematic approach to analyzing similarity measures. Without
going into details of specific application, we cannot objectively compare
measures. However, it is not reasonable to do the evaluation only with respect
to the final application. Consider for example the use of similarity measures
in recommendation of items in an learning system. Evaluating the quality of
recommendations is a complex task even if we compare just a single version of
the recommendation algorithm to a control group. It is not feasible to evaluate
variants of the recommendation algorithm for many similarity measures.

We thus need to analyze similarity measures even without considering specifics
of a particular application. Such evaluation cannot give us verdicts about
which measures are good or bad. But we can evaluate which decisions in the
similarity computations are really important -- which computation pipelines
lead to different results. In this way we can narrow the number of measures
that need to be explored for a particular application from hundreds to few
cases.

We propose to perform the evaluation on several levels of abstraction. All
these evaluations lead to results in the form of a matrix (visualized in the
form of heatmap in our figures). The interpretation of these matrices, however,
differs:
\begin{enumerate}
\item At the first level, we analyze the similarity of items with respect to
  specific measure (Figure~\ref{fig:similarity-examples}).
\item At the second level, we analyze agreement between different similarity
  measures (Figure~\ref{fig:robomise-corr-measures2}). Each cell of the matrix
  is now computed by comparing two similarity matrices from the first level of
  analysis.
\item At the third level, we analyze the impact of agreement measure
  (Figure~\ref{fig:corr-agreement}). Each cell of the matrix now corresponds to
  comparison of two agreement matrices from the second level of analysis.
\end{enumerate}

\subsection{Recommendations for Similarity Computation}

The suitable choice of a similarity measure depends, of course, on a particular
setting (programming environment, characteristics of available input data) and
the particular application of the measure. However, it is useful to have a basic
default choice which can serve as a baseline, which can be further improved.

Based on our explorations, we propose to use the following steps to compute
such a default similarity measure:
\begin{enumerate}
\item Use item solutions as input data.
\item Compute feature matrix based on the data using the basic bag-of-words
  approach -- computing number of occurrences of natural programming keywords.
\item Normalize the feature matrix, specifically using some variant of the
  TF-IDF transformation.
\item Compute item similarity based on the feature matrix using Euclidean
  distance of vectors in the normalized feature matrix.
\end{enumerate}

The aspect that can most importantly change the results of the computation is
the first step -- the choice of input data. We believe that item solutions are
good default, because for introductory programming problems they are basically
always available and the basic bag-of-word analysis can be easily performed in
different settings. For performance data to be useful, it is necessary to
collect large data on learner behavior, which limits their applicability. The
form of item statement data can depend on particular application (e.g., it
differs significantly between our robot programming problems and Python
programming problems), which makes its use more application specific.

\subsection{Limitations and Future Work}

A strong and unique aspect of our work is that we performed our exploration in
three different programming environments, which forced us to approach the
computation of similarity measures in a general way and allowed us to check
generality of the results. A limitation is the size of used item sets -- with
respect to real life applications these are still limited. For each environment
we used between 25 and 75 items. For realistic applications of similarity
measures in adaptive systems, a larger item set would be needed. Although we
believe that the larger item set should not significantly change the results,
it would be useful to explore similarity measures over large data sets.

We provide analysis of similarity measures mostly in application independent
way. As argued above, we believe that this is a necessary step. Based on the
results of the current analysis, it would be useful to analyze selected
similarity measures with respect to specific applications. Specific question
that requires attention is that whether different applications require
different similarity measures, or whether we can use one reasonably general
similarity measure.

In the current work we discuss only flat features -- each feature is treated
independently from others and all are treated on the same level. In
programming, however, features are typically interconnected and can be
naturally expressed using taxonomy or ontology. For example, we can have a
feature ``binary operator'' with subfeatures addition, multiplication,
division, etc.

In the processing of item statements we may want to distinguish superficial
similarity (related to the story or general topic of an item) and intrinsic
similarity (related to the way an item is solved). The basic bag-of-word model
that we discussed for processing item statements would not be able to
distinguish between these. For example in mathematics, word problems typically
contain significant story aspect and the basic use of bag-of-word model would
lead to dominance of the superficial similarity based on the story. In the
programming problems that we used the story aspects is not used, but in another
programming settings it may be more relevant. The superficial similarity may be
also useful for applications (e.g., we may want to present the learner a
sequence of problems with a similar story), but it should be treated separately
from the intrinsic similarity.


\balance{}
\bibliographystyle{SIGCHI-Reference-Format}
\bibliography{similarity-pp}

\end{document}